\pdfoutput=1  

\documentclass[letterpaper, 10 pt, conference]{ieeeconf}  

\IEEEoverridecommandlockouts                              

\usepackage{amsmath} 
\usepackage{amssymb}  
\usepackage{graphicx} 
\usepackage{microtype} 
\usepackage{url}
\usepackage{booktabs}
\usepackage{listings}
\title{\LARGE \bf
Neuro-Symbolic Participation Governance for \\Verifiable AI Agents in Open Digital Twin Ecosystems
}

\author{Juan Li$^{1}$, Wei Cai$^{2}$ and Yan Bai$^{2}$
\thanks{*This work was partially supported by National Science Foundation (NSF) with award number 2334197, 2218046, 1921576 and 2334196.}
\thanks{$^{1}$Juan Li is with Department of Computer Science at North Dakota State University, Fargo, ND, USA.
        {\tt\small J.Li@ndsu.edu}}%
\thanks{$^{2}$Wei Cai and Yan Bai are with the School of Engineering and Technology, University of Washington, Tacoma, WA, 98402, USA.
        {\tt\small \{weicaics, yanb\}@uw.edu }}%
\thanks{\copyright~2026 IEEE. Personal use of this material is permitted.
        Permission from IEEE must be obtained for all other uses, in any current
        or future media, including reprinting/republishing this material for
        advertising or promotional purposes, creating new collective works, for
        resale or redistribution to servers or lists, or reuse of any
        copyrighted component of this work in other works.}%
}

\usepackage{xcolor}
\newif\ifshowrevisions
\showrevisionstrue

\newcommand{\rev}[1]{\ifshowrevisions\textcolor{black}{#1}\else#1\fi}

\begin{document}

\maketitle
\thispagestyle{empty}
\pagestyle{empty}

\begin{abstract}
Autonomous AI agents, increasingly empowered by large language models, are becoming important components of human-machine systems for high-stakes decision support in digital twin ecosystems. However, existing multi-agent systems often lack robust verification for identity, capability, and policy compliance, especially in decentralized environments spanning multiple institutions. This paper proposes a neuro-symbolic decentralized governance framework for verifiable agents in collaborative digital twin environments. By representing agents through multi-layer semantic profiles, the framework bridges probabilistic neural reasoning with deterministic institutional governance, thereby supporting trustworthy human-AI collaboration and meaningful human oversight. Capabilities are grounded in formal domain ontologies to enable machine-interpretable, policy-aware, and context-sensitive participation. These credentials, issued by organizational authorities, are validated via blockchain-based smart contracts, ensuring auditable participation without exposing sensitive data. We demonstrate the framework using a decision-support prototype with clinic, digital twin, and wearable provider agents effectively prevents unauthorized interaction and enforces institutional policies with manageable overhead. Our findings suggest that neuro-symbolic decentralized governance provides a scalable and trustworthy pathway for safe human-machine collaboration across institutional boundaries.
\end{abstract}

\section{INTRODUCTION}
Autonomous AI agents, increasingly empowered by large language models (LLMs) capable of complex reasoning and tool interaction, are becoming integral to modern decision-support and human-machine systems~\cite{ferrag2026llmreasoningautonomousai}. In intelligent environments such as Digital Twin (DT) ecosystems~\cite{TRIPATHI2024107424}, these agents must collaborate across organizational boundaries to integrate complementary capabilities. For example, a DT agent for clinical simulation may rely on a Clinic Agent for real-time data and a Wearable Provider Agent for physiological metrics to produce predictive insights. Unlike conventional sensing agents, DT agents embed reasoning models that directly influence downstream outcomes, requiring rigorous guarantees of identity, semantic competence, and policy alignment. From a human-machine systems perspective, the core challenge is not only enabling autonomous coordination among agents, but also ensuring that human stakeholders can trust, supervise, and audit agent-mediated decisions in safety-critical workflows.

The full potential of collaborative AI agents is constrained by the decentralized nature of open environments. These ecosystems involve \textit{heterogeneous stakeholders} operating under distinct governance structures, the \textit{absence of a universal central authority} trusted to manage identities or enforce policies across institutions, and \textit{dynamic participation} as agents frequently enter and leave workflows. In high-stakes domains, especially healthcare, implicit trust and static access rules are inadequate. Unverified agents may introduce biased models or operate outside regulatory constraints, while centralized coordination creates single points of failure, reduces institutional autonomy, and limits transparency in collaborative decision-making. As multi-agent ecosystems grow, collaboration must rely on governance, not assumed trust, using verifiable identity, expertise, and policy compliance.

To address these limitations, we propose a neuro-symbolic decentralized governance framework for verifiable AI agents in collaborative DT ecosystems. It bridges the neural reasoning of modern autonomous agents with the symbolic requirements of formal governance. Each agent is represented through a multi-layer semantic profile that grounds probabilistic capabilities in formal domain ontologies (e.g., SNOMED CT), enabling machine-interpretable, policy-aware participation. These capabilities are encoded as cryptographically verifiable credentials (VCs) issued by institutional authorities and validated through a decentralized trust layer implemented with blockchain-based smart contracts. The proposed trust layer supports heterogeneous stakeholders through issuer-bound DIDs and VCs, decentralized on-chain validation, and dynamic participation via time-bounded credentials, revocation, and session logging. This approach authorizes and audits participation without exposing sensitive internal data, transforming collaboration into a structured and accountable process. Therefore, decentralized governance serves as the interface layer that keeps autonomous machine agents aligned with institutionally defined human objectives, responsibilities, and oversight requirements.

Our primary contributions are three-fold. First, we develop neuro-symbolic agent profiles that model agents through identity, functional capability, and operational constraints, thereby integrating high-level neural intelligence with symbolic guardrails. Second, we propose a knowledge-centric verification approach that grounds agent capabilities in domain-specific Knowledge Graphs and formal ontologies, enabling agents to prove semantic competence rather than merely asserting roles. Third, we implement a decentralized governance infrastructure based on a blockchain registry and policy engine that validates credentials and records interactions, providing a scalable and tamper-resistant audit trail for collaborative workflows.


\section{Related Work}

\subsection{Trust in Multi-Agent Systems}

Trust management in open multi-agent systems (MAS) has relied on reputation and behavioral models to estimate reliability through past interactions~\cite{ramchurn2004trust,huynh2006fire,yu2013survey}. While these approaches effectively identify malicious actors over time, they lack mechanisms to verify whether an agent possesses the semantic qualifications or institutional authorization required for a specific task. Recent large language model (LLM)-based agent frameworks, such as AutoGen~\cite{wu2023autogen}, demonstrate sophisticated autonomous coordination but offer limited guardrails for formal policy compliance or expertise verification~\cite{guo2024large}. Our work addresses this gap by shifting from behavioral estimation to credential-based participation governance.

\subsection{Decentralized Identity and Verifiable Credentials}

Decentralized identity provides a foundation for verifiable participation by enabling entities to prove attributes without centralized providers. The W3C standards for Decentralized Identifiers (DIDs)~\cite{w3c-did-core} and Verifiable Credentials (VCs)~\cite{w3c-vc-11} establish a cryptographic framework for cross-institutional trust. While frameworks like SAFE~\cite{safe2015} and various blockchain-based models~\cite{liu2023survey,9422738} have successfully implemented credential-based authorization in distributed systems, they typically focus on identity or static attributes. Our framework extends these standards by incorporating ontology-grounded capability subgraphs into the credential structure, enabling the verification of semantic competence alongside identity.

\subsection{Neuro-Symbolic AI and Knowledge-Grounded Agents}

The integration of neural learning with symbolic reasoning has emerged as a crucial paradigm for improving AI reliability and explainability~\cite{garcez2023neurosymbolic,hitzler2022neural}. Knowledge graphs and formal ontologies further support this integration by providing structured representations for complex domain reasoning~\cite{delong2024neurosymbolic}. However, most existing neuro-symbolic approaches focus on enhancing the internal reasoning of individual models rather than governing the collaborative interactions among multiple agents. In contrast, our framework employs neuro-symbolic semantic profiles as machine-verifiable representations to enforce participation constraints in open ecosystems.

\subsection{Digital Twin Ecosystems and Agent Collaboration}

Digital twin ecosystems increasingly rely on distributed agents to integrate heterogeneous data and predictive models across functional boundaries~\cite{katsoulakis2024digital,drummond2024definitions}. Prior research has explored agent-based coordination for healthcare digital twins~\cite{croatti2020integration} and proposed general multi-agent architectures for digital twin synchronization~\cite{10.1145/3697350}. Nevertheless, existing studies prioritize interoperability and data modeling over the verification of agent trustworthiness and expertise. We address this limitation by introducing a decentralized governance layer that ensures only authorized and semantically qualified agents can participate in collaborative digital twin workflows.

\section{Methodology}

\subsection{System Architecture}

\begin{figure}[htbp]
    \centering
  \includegraphics[width=\columnwidth]{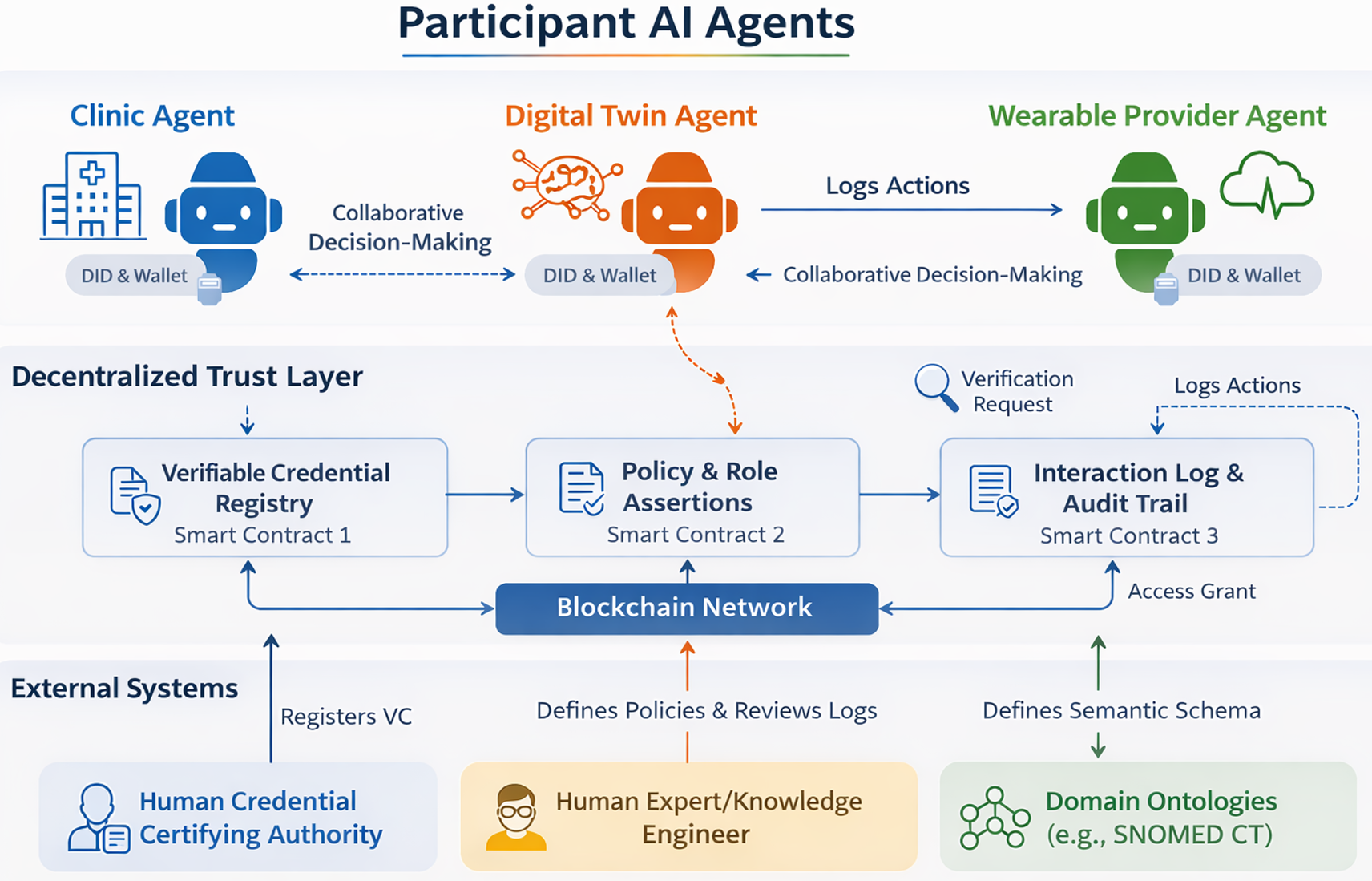}
    \caption{Decentralized governance architecture for verifiable multi-agent collaboration integrating AI agents, blockchain-based trust infrastructure, and human policy authorities.}
    \label{fig:agent-representation}
\end{figure}

The proposed framework enables verifiable multi-agent collaboration through a decentralized governance model that integrates human oversight with machine-verifiable participation. As illustrated in Fig.~\ref{fig:agent-representation}, the system consists of three interacting layers: the \textit{Participant AI Agents Layer}, where LLM-empowered agents collaborate on decision tasks; the \textit{Decentralized Trust Layer}, which manages identity and policy enforcement via blockchain; and the \textit{External Governance Layer}, where authorities define policies and maintain domain ontologies. We formalize a governed collaboration session as a quadruple $\mathcal{C} = <A, R, P, L>$, where $A$ is the set of participating agents, $R$ denotes assigned roles, $P$ represents the governing policies, and $L$ is the decentralized ledger for auditability. Participation is permitted only when an agent's credentials satisfy the specific task constraints, such that $\textit{Verify}(A_i, R_j, P_k) = \textit{True}$, \rev{where
\[
\begin{aligned}
\textit{Verify}(A_i, R_j, P_k) :=&\ \textit{ValidReg}(A_i) \wedge \textit{CapMatch}(A_i, R_j) \\
&\wedge \textit{PolicyMatch}(A_i, P_k).
\end{aligned}
\]
Here, $\textit{ValidReg}(A_i)$ checks issuer authorization, credential status, and temporal validity; $\textit{CapMatch}(A_i, R_j)$ checks whether the agent's ontology-grounded capability claim subsumes the role requirement; and $\textit{PolicyMatch}(A_i, P_k)$ checks that the submitted policy constraints equal the registered policy commitment and satisfy the task policy. It is also fail-closed under revocation because each authorization request and each interaction-log append re-evaluates registry validity on-chain. In deployment, the propagation bound is one new block plus the inclusion delay of the next transaction. We treat $\textit{Verify}$ as an admission-control predicate, not as a guarantee over the downstream clinical recommendation itself; output-level assurance is left to future work.}

\subsection{Verifiable Agent Profile and Credential Mapping}

Each agent $A_i$ is represented by a multi-layer neuro-symbolic profile $\mathcal{S}_i = \langle \mathcal{I}_i, \mathcal{C}_i, \mathcal{P}_i \rangle$ encoded as a W3C-compliant Verifiable Credential (VC). \rev{The profile is \emph{neuro-symbolic} because it binds learned or data-driven agent competence to explicit symbolic commitments that can be checked deterministically.} The identity layer $\mathcal{I}_i$ links the agent to its authority through a Decentralized Identifier (DID) and provenance chain, enabling on-chain origin verification and spoofing protection. The capability layer $\mathcal{C}_i$ captures the agent's neural or statistical function, such as prediction, simulation, or time-series interpretation, as a subgraph $\mathcal{G}_i \subseteq \mathcal{O}$ of a domain ontology $\mathcal{O}$ (e.g., SNOMED CT). For efficiency, a \texttt{capabilityHash} is computed as Keccak-256 over the canonicalized, sorted subgraph representation to provide a compact on-chain fingerprint. The policy layer $\mathcal{P}_i$ specifies symbolic governance constraints $\Phi_i$ (e.g., HIPAA compliance) and temporal bounds (\texttt{validFrom}/\texttt{validUntil}), which map to the smart contract's \texttt{expiresAt} validation and status checks for real-time participation control. A \texttt{policyHash} is anchored on-chain to verify compliance without exposing full subgraphs. \rev{The resulting VC therefore ties a learned agent's functional claim to ontology-grounded roles and institutional rules inside a single cryptographically signed object, as illustrated below.}

\noindent\textbf{Example VC Payload for a Clinic Agent}
\begin{lstlisting}[language={},basicstyle=\scriptsize\ttfamily]
{
 "@context": ["https://www.w3.org/2018/credentials/v1",
              "https://schema.org/AgentOntology"],
 "type": ["VerifiableCredential","AgentParticipationCredential"],
 "issuer": "did:ethr:0xHospitalAuthority...",
 "credentialSubject": {
   "id": "did:ethr:0xClinicAgent...",
   "identityLayer": {
     "organizationDID": "did:ethr:0xHospitalX...",
     "provenanceChain": ["TrainingRegistry:v2.1",
                         "ClinicalBoard:Approval:2024-03"]
   },
   "capabilityLayer": {
     "ontologyRef": "snomed:Diabetes_Risk_Prediction",
     "knowledgeSources": ["SNOMED-CT:73211009","LOINC:2345-7"],
     "capabilityHash": "keccak256(serialized_capability_subgraph)"
   },
   "policyLayer": {
     "constraints": [
       {"rule":"DataPrivacy","standard":"HIPAA",
        "status":"validated"},
       {"rule":"ConsentRequired","scope":"PatientData",
        "enforcement":"strict"},
       {"rule":"TemporalBound","validFrom":"2024-01-01",
        "validUntil":"2025-12-31"}
     ]
   }
 },
 "proof": {
   "type": "EcdsaSecp256k1Signature2019",
   "verificationMethod": "did:ethr:0xHospitalAuthority...#key-1",
   "proofValue": "0x..."
 }
}
\end{lstlisting}

\subsubsection{On-Chain vs. Off-Chain Management}

To preserve privacy while ensuring verifiability, the framework adopts a hybrid storage architecture. Sensitive agent data remains off-chain in distributed storage such as IPFS, including full VC payloads with detailed capability descriptions, policy specifications, raw knowledge graph subgraphs, training provenance records, and organizational metadata. Conversely, only cryptographic anchors and logic-critical pointers are registered on the smart contract registry. This on-chain data includes the agent DID-to-record mapping, containing Keccak-256 fingerprints of the canonicalized capability and policy subgraphs. To enable real-time participation control, the registry also maintains the credential lifecycle status (e.g., \texttt{Active}, \texttt{Suspended}, \texttt{Revoked}), temporal validity timestamps (\texttt{issuedAt} and \texttt{expiresAt}), and authorized issuer registry with associated revocation lists. This separation ensures that the blockchain serves as a verification anchor rather than a data store, allowing any verifier to confirm credential integrity without accessing the underlying sensitive data. For experimental efficiency, the on-chain fingerprints (\texttt{capabilityHash} and \texttt{policyHash}) are computed over canonicalized, sorted representations of the respective ontology class names extracted from the VC payload.

\subsubsection{Smart Contract Validation Logic}

The governance layer is implemented as a set of interconnected Solidity smart contracts deployed on an EVM-compatible blockchain. These contracts describe specific functions that process participation requests by comparing agent credentials against the symbolic policy nodes in the registry, as shown below:

\noindent\textbf{AgentRegistry Contract}

\begin{lstlisting}[language=Java,basicstyle=\scriptsize\ttfamily]
contract AgentRegistry {
  struct AgentRecord {
    bytes32 capabilityHash;  // H(G_i)
    bytes32 policyHash;      // H(Phi_i)
    address issuer; uint256 expiresAt;
    CredentialStatus status;
  }
  mapping(address => AgentRecord) public agents;
  mapping(address => bool) public authorizedIssuers;

  function registerAgent(address agentDID, bytes32 cap,
    bytes32 policy, uint256 expiry) external onlyAuthorizedIssuer {
    agents[agentDID] = AgentRecord({
      capabilityHash: cap, policyHash: policy,
      issuer: msg.sender, expiresAt: expiry,
      status: CredentialStatus.Active
    });
  }

  function verifyAgent(address agentDID) public view returns (bool) {
    AgentRecord memory record = agents[agentDID];
    return record.status == CredentialStatus.Active
      && block.timestamp <= record.expiresAt
      && authorizedIssuers[record.issuer];
  }

  function revokeCredential(address agentDID) external onlyAuthorizedIssuer { //... }
  function suspendCredential(address agentDID) external onlyAuthorizedIssuer { //... }
  function getAgentRecord(address agentDID) external view returns (AgentRecord memory) { //... }
}
\end{lstlisting}

\noindent\textbf{GovernancePolicy Contract}

\begin{lstlisting}[language=Java,basicstyle=\scriptsize\ttfamily]
contract GovernancePolicy {
  struct Policy {
    bytes32 requiredCapabilityHash;
    // ... [other task constraints]
  }
  // ... [state variables and policy registry]

  function authorizeRole(address agentDID, bytes32 taskId,
    bytes32 capHash, bytes32 policyHash, bytes32 roleId,
    bytes32 sessionToken, uint256 expiry) external {
    require(registry.verifyAgent(agentDID), "Identity invalid");

    AgentRegistry.AgentRecord memory record = registry.getAgentRecord(agentDID);
    require(capHash == record.capabilityHash, "Cap hash mismatch");
    require(policyHash == record.policyHash, "Policy hash mismatch");

    // ... [task-specific policy matching and binding]
    emit AccessGranted(agentDID, taskId, sessionToken, expiry);
  }
}
\end{lstlisting}

\textit{AgentRegistry} contract\footnote{Sepolia Contract Address: 0x2a01061c0ab8922b5e1283f053dbb7c63eb77a98} serves as the primary identity anchor, managing agent registration and providing real-time status and temporal validity verification. The \textit{GovernancePolicy} contract\footnote{Sepolia Contract Address: 0x38ac7f0d116850e3596eeeaeaf50120eb49bb22d} acts as the enforcement engine, ensuring that agents satisfy both semantic capability and institutional policy requirements before granting conditional access.

\subsection{Collaboration Governance Workflow}

\subsubsection{Participation Lifecycle}

The lifecycle of an agent joining a governed collaboration task consists of five sequential phases. Initially, in \textbf{Phase 1 (Participation Request)}, an agent submits its DID, the target task identifier, and the off-chain location (e.g., IPFS CID) of its full Verifiable Credential to the governance smart contract. Subsequently, in \textbf{Phase 2 (Credential Retrieval and Hash Verification)}, the trust layer retrieves the off-chain VC, computes a Keccak-256 hash over the payload, and compares it against the on-chain fingerprints in the \texttt{AgentRegistry} to ensure integrity. In \textbf{Phase 3 (Off-chain Capability Reasoning)}, the trust layer resolves the agent's full capability subgraph from the VC against the domain ontology using an OWL reasoner (e.g., HermiT). This enables flexible matching by verifying whether the agent's specific semantic skills subsume the required task roles, such as a Clinical Risk Assessor. Unlike rigid on-chain checks, this off-chain reasoning leverages the ontology's expressive power, while on-chain hashes remain solely for integrity verification. This is followed by \textbf{Phase 4 (Policy Compliance Verification)}, where the \texttt{policyHash} is verified against task requirements to enforce constraints such as HIPAA compliance or consent validation. Finally, in \textbf{Phase 5 (Conditional Access Grant)}, the smart contract emits an \texttt{AccessGranted} event and records authorization, issuing a time-bound session token for governed participation.

\subsubsection{Dynamic Role Binding}\label{sec:role}

Static role assignment is insufficient for open digital twin ecosystems where task requirements evolve. Our framework implements dynamic role binding through a multi-step inference process. When a task is instantiated, the coordinator queries the representation ontology to determine the required capability set (e.g., SNOMED CT concepts). The trust layer then retrieves the agent's ontology-grounded capabilities by resolving its \texttt{capabilityHash}. An OWL reasoner (e.g., HermiT) then infers the most specific task role that the agent's capability subgraph subsumes. For example, an agent covering cardiac risk assessment concepts is bound in real-time to the \textbf{Clinical Risk Assessor} role, while an agent with only heart rate monitoring capabilities is assigned a \textbf{Data Provider} role with restricted access. This role resolution occurs off-chain to minimize blockchain gas consumption, while the final binding is permanently recorded on-chain via the \texttt{GovernancePolicy} contract.

\subsubsection{Immutable Interaction Logging}

All agent interactions within a governed collaboration session are recorded in an append-only \textbf{Interaction Log} implemented as a smart contract\footnote{Sepolia Contract Address: 0x0d6f8f43c3394003b13ccc300c107c890e723038}. To ensure dynamic governance, each log entry requires a real-time validity check via \texttt{verifyAgent(agentDID)}, preventing revoked or expired agents from contributing further to the task. This immutable ledger provides three critical governance benefits. First, it ensures \textit{decision provenance}, as every prediction or action produced by an agent can be traced back to its specific inputs, bound roles, and verified credentials at the time of execution. Second, the framework supports \textit{post-hoc auditability}, enabling governance authorities to reconstruct decision chains and verify that each step complied with institutional policies. Finally, the blockchain provides inherent \textit{tamper resistance}, ensuring that interaction records cannot be retroactively altered, thereby maintaining long-term accountability across institutional boundaries. The combination of credential verification, dynamic role binding, and immutable logging ensures that collaboration in open digital twin ecosystems is governable, auditable, and accountable.

\section{Experimental Evaluation}

We evaluate the proposed framework in an ontology-grounded simulation involving three representative agents: a \textbf{Clinic Agent} (ClinicalDataAccess, PatientRecordQuery), a \textbf{Digital Twin Agent} (CardiacRiskAssessment, ECGInterpretation), and a \textbf{Wearable Agent} (HeartRateMonitoring, ActivityTracking). Each agent is governed by the neuro-symbolic profile and blockchain trust layer defined in Section III. The evaluation focuses on security resilience, dynamic policy enforcement, scalability, and system overhead.

\subsection{Experimental Setup}

The Agent Representation Ontology is implemented in OWL using Owlready2 and the HermiT reasoner, defining nine capability and four regulatory policy classes.

Each agent's capability and policy subgraphs are serialized, sorted, and hashed via Keccak-256 to produce compact on-chain fingerprints. \rev{The decentralized trust layer was evaluated locally on the Hardhat Network using Solidity v0.8.20, while the corresponding contracts were also deployed to the Sepolia testnet for public auditability.} We deployed three primary contracts: the \textit{AgentRegistry} for registration and revocation, the \textit{GovernancePolicy} for multi-stage role authorization, and the \textit{InteractionLog} for immutable recording with real-time validity checks. Additionally, a baseline \texttt{BaselineMAS}\footnote{Sepolia Contract Address: 0x549d33dbeae29a97908c73f405830d31d97ad111} contract with only address-based registration and no semantic or policy verification was deployed to represent the implicit trust models prevalent in existing MAS frameworks. \rev{The evaluation uses 42 executable test cases grouped into 14 security tests, 10 dynamic-policy tests, 10 scalability tests, and 8 overhead tests. Table~\ref{tab:attack-matrix} reports the 11 \emph{unique} attack scenarios covered by the security suite; the remaining three security tests are control or state-transition checks required to realize expiry, issuer-revocation propagation, and mid-session invalidation on-chain. We report admission outcome, revert reason correctness, false rejection rate, authorization latency, gas per function, and full-session overhead.} All 42 test cases were executed on a Linux workstation (AMD Ryzen, 32GB RAM). For cross-chain latency projections, block confirmation times were modeled based on published parameters for Ethereum L1 (12s), Optimism (2s), Base (2s), Arbitrum (0.26s), Polygon PoS (2s), and Solana (0.4s).

\subsection{Results and Security Analysis}

\subsubsection{Security Scenario: Attack Matrix}

To assess resilience, we tested 11 attack scenarios across identity forgery, capability escalation, and policy violations. 
These 11 cases were selected to cover the minimal but representative threat surface exposed by our architecture: credential lifecycle abuse, semantic capability tampering, issuer-side trust compromise, unauthorized task participation, and logging misuse, thereby spanning the main attack paths that a decentralized clinical DT workflow would face without making the evaluation artificially redundant.
As summarized in Table~\ref{tab:attack-matrix}, our framework achieved a 100\% detection rate, whereas the baseline accepted all malicious requests. \rev{This result should be interpreted as an executable check of the implemented safety invariants rather than as an end-to-end guarantee of clinical correctness: the mechanism proves that unauthorized or policy-incompatible agents cannot enter the governed workflow, but it does not by itself validate the quality of the model output produced by a legitimately admitted agent.} \rev{Crucially, the framework provides precise diagnostic revert reasons: ``Agent identity invalid'' for registry-validity failures (A1--A3, A9); ``Capability credential mismatch'' for submitted capability claims that conflict with the registered credential (A4, A6--A8); ``Policy credential mismatch'' for a submitted policy hash that conflicts with the registered credential (A5); ``Policy constraint mismatch'' when an agent has the required capability but does not satisfy the task policy (A8b); and ``Capability mismatch'' when an agent's registered capability does not satisfy the task requirement (A10). This diagnostic capability supports post-hoc forensic analysis of rejected participation attempts.}

\begin{table}[htbp]
\centering
\caption{Attack Matrix: Detection Results}
\label{tab:attack-matrix}
\footnotesize
\renewcommand{\arraystretch}{0.85}
\begin{tabular*}{\columnwidth}{@{\extracolsep{\fill}}llcc@{\extracolsep{\fill}}}
\toprule
\textbf{ID} & \textbf{Attack Scenario} & \textbf{Proposed} & \textbf{Baseline} \\
\midrule
A1 & Unregistered agent & Reject & Accept \\
A2 & Expired credential & Reject & Accept \\
A3 & Revoked issuer & Reject & Accept \\
A4 & Capability impersonation & Reject & Accept \\
A5 & Partial capability match & Reject & Accept \\
A6 & Tampered off-chain VC & Reject & Accept \\
A7 & Replayed old credential & Reject & Accept \\
A8 & Wrong capability \& policy & Reject & Accept \\
A8b & Correct cap., wrong policy & Reject & Accept \\
A9 & Temporal expiry mid-session & Reject & Accept \\
A10 & Combined cap. + policy & Reject & Accept \\
\midrule
\multicolumn{2}{l}{\textbf{Detection Rate}} & \textbf{100\%} & \textbf{0\%} \\
\bottomrule
\end{tabular*}
\end{table}

\subsubsection{Dynamic Policy Adaptation}

Real-time governance was verified along four dimensions. First, for \textbf{policy hot updates}, the Digital Twin Agent was initially authorized under HIPAA, PatientConsent, and GDPR at $T_1$; after the policy was upgraded to include \texttt{DataAnonymization}, requests at $T_2$ were immediately rejected, confirming that changes take effect in the next block. Second, \textbf{credential revocation propagation} was validated by revoking one agent via \texttt{revokeCredential()}, after which subsequent attempts to record interactions in the \texttt{InteractionLog} were rejected with \rev{``Agent no longer valid.''} Removing an authority from the \texttt{authorizedIssuers} registry likewise propagated to all agents it had issued, confirming cascading revocation. \rev{Under the current architecture, the worst-case propagation delay is bounded by one newly mined block plus the inclusion of the next authorization or logging transaction.} Third, \textbf{cross-task policy isolation} ensured that an agent authorized for Task 1 (diabetes risk assessment) could not participate in Task 2 (cardiac monitoring), even when both tasks coexisted, thereby enforcing capability-based task boundaries. Finally, across 100 consecutive authorization requests from a legitimate agent, the framework yielded a \textbf{zero false rejection rate} (0.0\%). Interaction logs recorded before a policy update also remained intact and readable, confirming that enforcement is non-retroactive and preserves audit integrity.

\subsubsection{Scalability}

We measured authorization performance with up to 50 agents. 
We capped the experiment at 50 agents because this scale realistically reflects cross-institution DT collaboration in our prototype while still permitting repeated end-to-end runs with full logging and policy checks under controlled local conditions. As shown in Table~\ref{tab:scalability}, gas consumption remains constant at $\sim$63,000 per call, confirming $O(1)$ verification complexity relative to ecosystem size. \rev{Per-agent authorization latency, computed as total batch time divided by the number of agents, remained approximately constant over the tested range, with small measurement variation and zero observed failures.}

\begin{table}[htbp]
\centering
\caption{Scalability: Authorization Performance}
\label{tab:scalability}
\footnotesize
\renewcommand{\arraystretch}{0.85}
\begin{tabular*}{\columnwidth}{@{\extracolsep{\fill}}ccccc@{\extracolsep{\fill}}}
\toprule
\textbf{Agents} & \textbf{Total (ms)} & \textbf{Avg/Agent (ms)} & \textbf{Avg Gas} & \textbf{Failures} \\
\midrule
10 & 9 & \rev{0.9} & 63,051 & 0 \\
20 & 15 & \rev{0.8} & 63,050 & 0 \\
30 & 28 & \rev{0.9} & 63,049 & 0 \\
40 & 35 & \rev{0.9} & 63,050 & 0 \\
50 & 34 & \rev{0.7} & 63,049 & 0 \\
\bottomrule
\end{tabular*}
\end{table}

\subsubsection{Overhead Analysis}

Table~\ref{tab:gas} details the gas costs for individual operations. While the framework consumes more gas than the baseline MAS (total gas: 219,726; time: 3~ms for registration and access only) due to the 30 interaction log entries required for auditability (89.5\% of total gas), the core registration and authorization operations remain efficient (10.5\% of total gas). Table~\ref{tab:session} summarizes the overhead for a complete three-agent collaboration session.

\begin{table}[htbp]
\centering
\caption{Per-Function Gas Costs (50 runs)}
\label{tab:gas}
\footnotesize
\renewcommand{\arraystretch}{0.85}
\begin{tabular*}{\columnwidth}{@{\extracolsep{\fill}}lrrrl@{\extracolsep{\fill}}}
\toprule
\textbf{Operation} & \textbf{Avg Gas} & \textbf{Min} & \textbf{Max} & \textbf{Note} \\
\midrule
registerAgent & 182,813 & 182,802 & 182,814 & One-time \\
verifyAgent & 39,239 & 39,239 & 39,239 & View (free) \\
authorizeRole & 63,038 & 63,027 & 63,039 & Per task \\
recordInteraction & 209,693 & 209,332 & 226,444 & Per entry \\
revokeCredential & 32,411 & 32,388 & 32,412 & On demand \\
\bottomrule
\end{tabular*}
\end{table}

\begin{table}[htbp]
\centering
\caption{Full Collaboration Session Overhead}
\label{tab:session}
\footnotesize
\renewcommand{\arraystretch}{0.85}
\begin{tabular*}{\columnwidth}{@{\extracolsep{\fill}}lrr@{\extracolsep{\fill}}}
\toprule
\textbf{Phase} & \textbf{Time (ms)} & \textbf{Gas} \\
\midrule
Registration (3 agents) & 2 & 548,430 \\
Authorization (3 agents) & 3 & 189,129 \\
Interaction Logging (30 entries) & 23 & 6,297,900 \\
\midrule
\textbf{Total} & \textbf{28} & \textbf{7,035,459} \\
\bottomrule
\end{tabular*}
\end{table}

\rev{Table~\ref{tab:crosschain} provides an illustrative cross-chain projection using the locally measured gas consumption of one \texttt{authorizeRole} transaction (63,039 gas) and representative nominal block intervals and transaction-cost estimates. The final column gives a conservative upper bound on authorization latency, assuming all 50 independent authorization transactions are included one per block. In practice, latency is often lower because multiple transactions can be included in the same block, depending on gas limits, mempool conditions, and confirmation policies. Cost estimates are indicative and vary with network conditions. The Solana row is a deployment projection requiring a Solana-compatible implementation of the workflow.} \rev{Overall, the evaluations show that neuro-symbolic decentralized governance provides comprehensive participation security and dynamic policy enforcement with practical overhead, while cross-chain projections indicate feasible deployment costs and latency across the modeled platforms.}

\begin{table}[htbp]
\centering
\caption{\rev{Cross-Chain Cost Projection and Conservative Sequential Authorization-Latency Bounds}}
\label{tab:crosschain}
\footnotesize
\renewcommand{\arraystretch}{0.85}
\begin{tabular*}{\columnwidth}{@{\extracolsep{\fill}}lrrr@{\extracolsep{\fill}}}
\toprule
\textbf{Chain} & \textbf{Block (s)} & \textbf{Cost/tx (USD)} & \rev{\textbf{Serial Bound (s)}} \\
\midrule
Ethereum L1 & 12.0 & \$0.0029 & 600.0 \\
Optimism & 2.0 & \$0.00001 & 100.0 \\
Base & 2.0 & \$0.000007 & 100.0 \\
Arbitrum & 0.26 & \$0.00001 & 13.0 \\
Polygon PoS & 2.0 & \$0.000001 & 100.0 \\
Solana (proj.) & 0.4 & \$0.00025 & 20.0 \\
\bottomrule
\end{tabular*}
\end{table}

\section{Conclusion}

This paper presented a neuro-symbolic decentralized governance framework for verifiable AI agent collaboration in open digital twin ecosystems. By bridging the probabilistic reasoning of modern LLMs with the deterministic guardrails of symbolic ontologies and blockchain-enforced credentials, the proposed framework shifts multi-agent collaboration from assumption-based trust to a governance-driven, evidence-based process. Experimental evaluation across diverse security scenarios shows that it prevents unauthorized participation and enforces complex institutional policies with manageable overhead, supporting a scalable foundation for the emerging Internet of Agents. More broadly, the framework contributes to human-machine systems research by showing that trustworthy human oversight can be maintained when decision support is delegated to interacting autonomous agents across institutional boundaries. \rev{The present scope is admission governance and provenance rather than runtime verification of the clinical recommendation itself; future work will combine contract-level invariant checking with output monitoring and privacy-preserving assurance mechanisms.}


\section*{Code Availability}

The VeriGov-AI implementation, ontology artifacts, and experiment scripts are openly available at \url{https://github.com/weicaiuw/verigov-ai} (DOI: \url{https://doi.org/10.5281/zenodo.21706699}).

\bibliographystyle{ieeetr}
\bibliography{citation}

\end{document}